\documentclass[%
 pre,
 superscriptaddress,
 reprint,
 amsmath,amssymb,
 aps,
 floatfix
]{revtex4-2}

\usepackage{physics}
\usepackage{graphicx}
\usepackage{dcolumn}
\usepackage{bm}

\begin{document}

\title{A replica theory for the dynamic glass transition of hardspheres with continuous polydispersity} 

\author{Hyonggi Kim}
\affiliation{Graduate School of Arts and Sciences, Meguro-ku, University of Tokyo, Tokyo 153-8902, Japan}
\author{Atsushi Ikeda}
\affiliation{Graduate School of Arts and Sciences, Meguro-ku, University of Tokyo, Tokyo 153-8902, Japan}
\affiliation{Research Center for Complex Systems Biology, Universal Biology Institute, The University of Tokyo, Komaba, Tokyo 153-8902, Japan}

\date{\today}

\begin{abstract}
Glassy soft matter is often continuously polydisperse, in which the sizes or various properties of the constituent particles are distributed continuously. 
However, most of the microscopic theories of the glass transition focus on the monodisperse particles. 
Here, we developed a replica theory for the dynamic glass transition of continuously polydisperse hardspheres. 
We focused on the limit of infinite spatial dimension, where replica theory becomes exact. 
In theory, the cage size $A$, which plays the role of an order parameter, appears to depend on the particle size $\sigma$, and thus, the effective free energy, the so-called Franz-Parisi potential, is a functional of $A(\sigma)$. 
We applied this theory to two fundamental systems: a nearly monodisperse system and an exponential distribution system. 
We found that dynamic decoupling occurs in both cases; the critical particle size $\sigma^{\ast}$ emerges, and larger particles with $\sigma \geq \sigma^{\ast}$ vitrify, while smaller particles $\sigma < \sigma^{\ast}$ remain mobile. 
Moreover, the cage size $A(\sigma)$ exhibits a critical behavior at $\sigma \simeq \sigma^{\ast}$, originating from spinodal instability of $\sigma^{\ast}$-sized particles. 
We discuss the implications of these results for finite dimensional systems. 
\end{abstract}

\maketitle

\section{Introduction}

Soft matter, such as colloidal dispersions and emulsions, is often continuously polydisperse.
This type of system is not composed of a single type of particle, but the sizes or various properties of the constituent particles are distributed continuously.
Polydispersity is essential because it strongly affects the phase behavior of the system.
For example, polydispersity greatly complicates the phase diagram via the fragmentation process~\cite{Bareigts2020,Sollich2002}, and its small change dramatically alters the rate of the crystallization~\cite{Auer2001,Schope2006}.
Understanding and predicting the phase behavior of continuously polydisperse systems remain challenges in soft matter and statistical physics~\cite{Sollich2002}.
Theoretically, one of the difficulties is that an infinite number of order parameters are generally required since an infinite number of types of particles are present in a system.

Vitrification is a common phenomenon in dense polydisperse particles~\cite{Pusey1986}.
Because polydispersity inhibits crystallization, these systems often remain in a metastable liquid state when overcompressed.
Then, at sufficiently high densities, each particle is caged, i.e., confined by the surrounding particles, and the system undergoes a glass transition~\cite{DebenedettiBook}.
The glass transition of polydisperse particles has long been studied in experimental and simulation studies~\cite{Brambilla2009,Bernu1987,Kob1995}.
In particular, in recent simulations, continuously polydisperse systems have attracted considerable attention because highly effective numerical algorithms have been developed~\cite{Ninarello2017}.
However, the impact of polydispersity on the glass transition has yet to be fully elucidated.
Existing experimental and simulation studies reported the following characteristics of the glass transition of polydisperse particles. 
When the polydispersity is extremely small, the phenomenology of the glass transition is almost the same as that of the monodisperse particles~\cite{Henderson1996,Zaccarelli2009}.
However, as the polydispersity increases, the glass transition density increases, or equivalently, the liquid state remains stable up to higher densities~\cite{Behera2017,Zaccarelli2015}.
Concomitantly, the relaxation dynamics of the particles start to depend on the particle size; smaller particles tend to have larger cage sizes and shorter relaxation times~\cite{Zaccarelli2015,Imhof1995,Hendricks2015,Heckendorf2017,Moreno2006,Voigtmann2009,Vaibhav2022,Pihlajamaa2023}.
When the polydispersity is large enough, dynamic decoupling occurs; larger particles vitrify, while smaller particles remain in a liquid state~\cite{Zaccarelli2015,Imhof1995,Hendricks2015,Moreno2006,Voigtmann2009,Vaibhav2022}.

From the theoretical side, microscopic theories of the glass transitions of polydisperse particles have yet to be developed.
The replica liquid theory (RLT)~\cite{ParisiBook} and mode coupling theory (MCT)~\cite{GotzeBook} are two basic microscopic theories of glass transitions.
The RLT was initially formulated for monodisperse particles~\cite{Mezard1999,ParisiReview}, and then extended to bidisperse particles~\cite{Coluzzi1999,Biazzo2009,Ikeda2016,Ikeda2017,Ikeda2019,Ikeda2021}.
In particular, a recent theory predicted the dynamic decoupling between large and small particles in bidisperse particles~\cite{Ikeda2021}.
However, there has been no attempt to extend the RLT to continuously polydisperse particles.
The status of the MCT is not far from that of the RLT.
The MCT for bidisperse particles has been developed, which predicts dynamic decoupling~\cite{Bosse1995,Voigtmann2011}.
However, the extension of the MCT to continuously polydisperse particles has been attempted only very recently~\cite{Laudicina2023}.

This study aimed to construct an RLT for continuously polydisperse particles.
The RLT is a thermodynamic theory of the glass transition, where the cage size $A$, the spatial extent of the cage, plays the role of an order parameter of the glass transition~\cite{ParisiBook}.
Then, the effective free energy $V$, the so-called Franz-Parisi (FP) potential, is calculated as a function of the cage size $A$, and the glass transition is predicted from the local and global minima of $V(A)$.
With increasing density or decreasing temperature, a local minimum of $V(A)$ emerges at some point. 
This describes the dynamic glass transition, which conceptually corresponds to the MCT transition~\cite{ParisiBook,Ikeda2010}.
In this study, we develop an RLT to describe the dynamic glass transition of continuously polydisperse particles.

We focus on hard spheres with continuous polydispersity in the infinite spatial dimension $d \to \infty$, where the RLT becomes exact. 
In Sec. II, we formulate a replica theory of continuously polydisperse hard spheres. 
We introduced the cage size $A$ as a function of the particle size $\sigma$, which corresponds to an infinite number of order parameters in continuously polydisperse systems.
As a result, the FP potential becomes a functional of the order parameter function, $V[A(\sigma)]$. 
By imposing the stationary point condition on $V[A(\sigma)]$, we derived a self-consistent equation of $A(\sigma)$, which describes the dynamic glass transition of continuously polydisperse hardspheres. 
In Sec.~III, we apply this formulation to a nearly monodisperse system. 
We show that this model exhibits dynamic decoupling; a critical particle size $\sigma^{\ast}$ emerges, and larger particles $\sigma \geq \sigma^{\ast}$ become glassy, while smaller particles $\sigma < \sigma*$ remain in a liquid state.
Moreover, the cage size $A(\sigma)$ exhibits a characteristic square-root singularity at $\sigma \simeq \sigma^{\ast}$, indicating that $\sigma^{\ast}$-sized particles are on the verge of spinodal instability. 
In Sec.~IV, we apply our formulation to an exponential distribution system, which is an infinite dimensional counterpart of the power-law distribution system. 
We found that this model has three phases: the liquid, glass, and partial glass phases. 
In the partial glass phase, dynamic decoupling occurs, and the cage size follows the same critical laws as in nearly monodisperse systems. 
Sec.~V summarizes our findings and discusses their implications for the glass transition of continuously polydisperse systems in finite dimensions.

\section{Theory}

\subsection{Model}
This study considers hard spheres with continuous polydispersity in the limit of spatial dimension infinity $d \to \infty$.
The number of particles in the system is denoted by $N$, and the volume is $V_{\rm box}$.
The total energy of the hard spheres is given by $U=\sum_{i<j} v(r_{ij}:\sigma_i,\sigma_j)$, where the interparticle interaction is
\begin{equation}
v(r_{ij}:\sigma_{i},\sigma_{j}) =  
\begin{cases}
  \infty & \text{if $r_{ij}\leq \sigma_{ij}$,} \\
  0       & \text{if $r_{ij}>\sigma_{ij}$.}
\end{cases}
\end{equation}
$\vb*{r}_{i}$ and $\sigma_i$ denote the position and diameter of the particle $i$, respectively, $r_{ij}=|\vb*{r}_{i}-\vb*{r}_{j}|$ is the distance between the particles $i,j$, and $\sigma_{ij}=(\sigma_{i}+\sigma_{j})/2$ is the mean particle diameter.

The control parameters of the system are the packing fraction $\phi$ and the probability distribution of the particle size $\varphi (\sigma )$.
However, to obtain sensible results in infinite dimensions, we need to work with scaled versions of these parameters, $\overline{\phi}$ and $\hat{\varphi}(\hat{\sigma})$. 
These parameters are introduced as follows. 
We first reparametrize the particle size as~\cite{Ikeda2021}
\begin{equation}
\sigma = \sigma_{0} \left(1+\frac{\hat{\sigma}}{d}\right),  
\label{sizescaling}
\end{equation}
where $\hat{\sigma}$ parametrizes the particle diameter in the infinite dimensional system and $\hat{\sigma}>0$ means that $\sigma$ is $O(1/d)$ larger than $\sigma_{0}$. 
Note that, without this reparametrization, the volume ratio between differently sized particles diverges at $d \to \infty$. 
Accordingly, we introduce the probability distribution of $\hat{\sigma}$ as $\hat{\varphi}(\hat{\sigma}) = \varphi (\sigma) \sigma_0/d$. 
We also introduce the scaled packing fraction $\overline{\phi}= 2^{d} \phi /d$ because the dynamic glass transition density depends on the dimension as $\phi_{\rm d} \propto 2^{-d} d$~\cite{ParisiBook}. 
Then, the scaled packing fraction $\overline{\phi}$ and particle size distribution $\hat{\varphi}(\hat{\sigma})$ are related as 
\begin{equation} \label{packingfraction}
\overline{\phi} = \overline{\phi}_{0} \int_{-\infty}^{\infty} d\hat{\sigma} \hat{\varphi}(\hat{\sigma}) e^{\hat{\sigma}}
\end{equation}
where
\begin{equation}
\overline{\phi}_{0} = \frac{N}{V_{\rm box}} \frac{d \pi^{d/2} \sigma_0^d}{\Gamma (1 + d/2 )}
\end{equation}
is the scaled packing fraction of the corresponding monodisperse system with a particle diameter $\sigma_0$.

\subsection{Equilibrium Liquid Theory}

We first consider the equilibrium liquid theory of this system.
The grand partition function of the continuously polydisperse liquid is given by~\cite{Wilding2002}
\begin{equation}
\varXi = \sum_{N=0}^{\infty} \frac{1}{N! \lambda^{Nd}} \int d\vb*{r}_{1}...d\vb*{r}_{N} \int d\sigma_{1}...d\sigma_{N} e^{\beta \sum_{i=1}^{N}\mu (\sigma_{i} )} e^{- \beta U} 
\label{partition}
\end{equation}
where $\beta$ is the inverse temperature and $\lambda$ is the thermal de Broglie wavelength. 
The polydispersity is taken into account through the particle size dependence of the chemical potential, $\mu ( \sigma )$. 

We introduce the density field for the polydisperse system as  
\begin{eqnarray}
\rho(\vb*{r},\sigma)= \left\langle \sum_{i=1}^{N}\delta (\vb*{r}-\vb*{r}_{i}) \delta (\sigma-\sigma_{i}) \right\rangle, 
\end{eqnarray}
which expresses the probability that $\sigma$-sized particles exist at the position $\vb*{r}$. 
We then used $\rho(\vb*{r},\sigma)$ to perform the virial expansion of the grand potential, following the standard method for the monodisperse systems~\cite{Morita1961,Stell}.   
Due to the same reasoning as for the monodisperse system~\cite{ParisiBook}, the higher-order terms $O(\rho^3)$ vanish in the $d \to \infty$ limit, and we obtain the free energy of the system:
\begin{align} \label{freeliq}
 & -\beta F = \int d{\vb*{r}} d\sigma \rho ({\vb*{r}},\sigma) (1- \log{\rho ({\vb*{r}}, \sigma )}+\log{\lambda^{d}} ) \nonumber \\
  & \quad +\frac{1}{2}\int d {\vb*{r}_{1}}  d {\vb*{r}_{2}} d\sigma_{1} d\sigma_{2} \rho ({\vb*{r}_{1}},\sigma_{1} ) \rho ({\vb*{r}_{2}},\sigma_{2} ) f (r_{12}, \sigma_{12} ), 
\end{align}
where $f (r_{12},\sigma_{12} )= e^{-\beta v (r_{12}:\sigma_{1},\sigma_{2} )}-1$ is the Mayer function.
We note that when the system is spatially uniform, the density field obeys $\rho (\vb*{r},\sigma ) = \frac{N}{V_{\rm box}} \varphi (\sigma )$.

\subsection{Replica Liquid Theory}

We now construct a replica theory for this system.
This study focuses on replica symmetric solutions since we focus on the dynamic glass transition.

For each $\sigma$-sized particle in the system, we add $s$ replica particles with a particle size $\sigma$ and consider a ``replica molecule'' composed of $s+1$ particles~\cite{ParisiBook}.
We then consider a liquid state theory for replica molecules.
The density field of this system is represented as  
\begin{align}
\rho (\overline{r},\sigma )= \left\langle \sum_{i} \prod_{a} \delta ({\vb*{r}}^{a}-{\vb*{r}}_{i}^{a} ) \delta (\sigma-\sigma_{i}) \right\rangle 
\end{align}
where $a \in [1,2,..,s+1]$ is the index of the replica and $\vb*{r}_{i}^{a}$ is the position of the $a$-th replica particle in the $i$-th replica molecule.
Here, we introduced a compact notation $\overline{r}=(\vb*{r}^{1},\vb*{r}^{2},...,\vb*{r}^{s+1})$. 
The density field $\rho (\overline{r},\sigma )$ represents the probability that replica molecules of $\sigma$-sized particles exist at position $\overline{r}$. 

\begin{widetext}
Using this density field, we can evaluate the free energy of the replica molecular liquid.
Because the procedure is analogous to that for the monodisperse case~\cite{ParisiBook}, we only show the outline.
We first perform the virial expansion of the grand potential and obtain the free energy of the replica molecular liquid as
\begin{align}
-\beta F_{s+1} = \int d\overline{x} d\sigma  \rho (\overline{x},\sigma) (1-\log{\rho (\overline{x},\sigma)})
+ \frac{1}{2} \int d\overline{x} d\overline{y} d\sigma_{1} d\sigma_{2} \rho (\overline{x},\sigma_1) \rho (\overline{y},\sigma_2) G(\overline{x},\overline{y},\sigma_{1},\sigma_{2}),  \label{Fs1}
\end{align}
where $\overline{x} = (\vb*{x}^{1},\vb*{x}^{2},...,\vb*{x}^{s+1})$ is the spatial coordinate expressing the position of a replica molecule and $G(\overline{x},\overline{y},\sigma_{1},\sigma_{2}) = \prod_{a=1}^{s+1} \exp{\left(-\beta v\left(|\vb*{x}^a-\vb*{y}^a|:\sigma_{1},\sigma_{2}\right)\right)} -1$ is the quantity corresponding to the Mayer function.
We now introduce the Gaussian ansatz for the density field~\cite{ParisiReview}: 
\begin{equation}
  \rho(\overline{x},\sigma) = \frac{N}{V_{\rm box}} \varphi (\sigma) \int dX (2\pi A(\sigma))^{-\frac{d}{2}} \exp \left(-\frac{1}{2A(\sigma)}\sum_{a=1}^{s+1}(x_{a}-X)^{2} \right) . \label{ansatz}
\end{equation}
This is the assumption that the particles in a replica molecule are distributed according to a Gaussian distribution with variance $A(\sigma)$, which is justified at $d \to \infty$~\cite{ParisiBook}.
The variance $A(\sigma)$ expresses the spatial extent of a replica molecule, corresponding to the cage size in a glass state.
We then substitute the ansatz Eq.~(\ref{ansatz}) into Eq.~(\ref{Fs1}), evaluate the integrals at $d \to \infty$, and calculate the FP potential $V$ using the formula $V = \lim_{s \to 0} \partial_s F_{s+1}$. 
With the introduction of the scaled cage size $\hat{A} (\hat{\sigma})= d^{2} A(\sigma) / \sigma_{0}^{2}$, the final result is 
\begin{align} 
& V\left[\hat{A}(\hat{\sigma})\right] = \frac{d}{2} \int d\hat{\sigma}_{1} \hat{\varphi} (\hat{\sigma}_{1}) \log{2\pi \hat{A}(\hat{\sigma}_{1})} \nonumber \\
&+ \frac{d}{2} \overline{\phi}_{0} \int d\hat{\sigma}_{1} d\hat{\sigma}_{2} \hat{\varphi} (\hat{\sigma}_{1} ) \hat{\varphi} (\hat{\sigma}_{2}) e^{\frac{\hat{\sigma}_{1}+\hat{\sigma}_{2}}{2}} \sqrt{2(\hat{A}(\hat{\sigma}_{1})+\hat{A}(\hat{\sigma}_{2}))} e^{-\frac{\hat{A}(\hat{\sigma}_{1})+\hat{A}(\hat{\sigma}_{2})}{2}} \int_{-\infty}^{\infty} dz e^{\sqrt{2(\hat{A}(\hat{\sigma}_{1})+\hat{A}(\hat{\sigma}_{2}))}z} \varTheta (z) \log{\varTheta(z)}, \label{FP}
\end{align}
where $\varTheta(z)=\frac{1}{2}(1+ \erf(z))$. 
Notably, the cage size $\hat{A}$ is a function of the particle size $\hat{\sigma}$ and the FP potential is a functional of $\hat{A}(\hat{\sigma})$, which reflects that the cage size generally depends on the particle size in polydisperse systems.
The factor $e^{(\hat{\sigma}_{1}+\hat{\sigma}_{2})/2}$ comes from the excluded volume of two particles with different particle sizes: $((\sigma_1 + \sigma_2)/2)^d$ in the $d \to \infty$ limit.
\end{widetext}

By imposing the stationary point condition on the FP potential (Eq.~\eqref{FP}), we obtain a self-consistent equation for the cage size.
In the polydisperse case, this calculation becomes a functional differentiation, $0 = \frac{\delta V}{\delta \hat{A}}(\hat{\sigma})$, which results in  
\begin{equation}
  \frac{1}{\hat{A}(\hat{\sigma})}= \overline{\phi}_{0} e^{\hat{\sigma}/2} \int d \hat{\sigma}' e^{\hat{\sigma}'/2} \hat{\varphi}(\hat{\sigma}') g\left(\frac{\hat{A}(\hat{\sigma})+\hat{A}(\hat{\sigma}')}{2}\right),  \label{determiningA}
\end{equation}
where 
\begin{equation}
g(A)= -\frac{\partial}{\partial A} \sqrt{4 A} e^{-A} \int_{-\infty}^{\infty} dz e^{\sqrt{4A}z} \varTheta (z) \log{\varTheta (z)}. 
\end{equation}
is the auxiliary function, as in the monodisperse case~\cite{ParisiBook}.
The self-consistent equation Eq.~(\ref{determiningA}) has a simple physical interpretation. 
The left-hand side comes from the ideal gas term in the FP potential, which favors increasing the cage size.
The right-hand side comes from the two-body interaction term in the FP potential, which penalizes the increase in the cage size.
The integration over $\hat{\sigma}'$ takes into account that two-body collisions with particles of various sizes occur. 

\subsection{Summary}

Thus far, we have constructed a replica theory for continuously polydisperse hard spheres.
The final result is the self-consistent equation Eq.~(\ref{determiningA}).
Here, we summarize our strategy of using this equation to determine the glass transition of the system.

The control parameters of the continuously polydisperse hard spheres in the infinite dimension are the scaled packing fraction $\overline{\phi}$ and the scaled particle size distribution $\hat{\varphi}(\hat{\sigma})$.
The packing fraction $\overline{\phi}$ is related to $\overline{\phi}_{0}$ by Eq.~(\ref{packingfraction}).
Therefore, once the system is specified by $\overline{\phi}$ and $\hat{\varphi}\left(\hat{\sigma}\right)$, Eq.~(\ref{determiningA}) is an integral equation for the unknown scaled cage size $\hat{A}(\hat{\sigma})$. 
The solution describes the dynamic glass transition.
The $\hat{\sigma}$-sized particles are in a liquid state if $\hat{A}(\hat{\sigma})$ diverges, while they are in a glass state if $\hat{A}(\hat{\sigma})$ is finite.
In the subsequent sections, we apply this strategy to two fundamental systems.

\section{Nearly monodisperse system}

In this section, we study the simplest continuous polydisperse hard spheres of the nearly monodisperse (NM) system. 
This system is defined by the particle size distribution: 
\begin{equation}
\hat{\varphi} (\hat{\sigma} ) = 
\begin{cases}
(1 -  \epsilon) \delta (\hat{\sigma} ) + \frac{\epsilon}{2\hat{\sigma}_{\rm M}} & \text{if $-\hat{\sigma}_{\rm M} \leq \hat{\sigma} \leq \hat{\sigma}_{\rm M}$,} \\
0   & \text{otherwise.} 
\end{cases}
\label{del_dist}
\end{equation}
The parameter $\epsilon$ describes the fraction of particles with $\hat{\sigma} \neq 0$, and $\hat{\sigma}_{\rm M}$ is the cutoff particle size.
We study the dynamic glass transition of this model in the limit of $\epsilon \to 0$ with sufficiently large $\hat{\sigma}_{\rm M}$, in which particles with $\hat{\sigma}\neq 0$ are extremely rare, and the system is nearly monodisperse.
In this limit, the packing fraction of this system satisfies $\overline{\phi} = \overline{\phi}_0$ (see Eq.~(\ref{packingfraction})). 

Hereafter, the cage size is denoted by $\hat{A} (\hat{\sigma},\overline{\phi})$ to emphasize its dependence on the particle size $\hat{\sigma}$ and the packing fraction $\overline{\phi}$. 
We substitute Eq.~(\ref{del_dist}) into Eq.~(\ref{determiningA}) and focus on the leading order of $\epsilon$. 
This is just the substitution of $\hat{\varphi} (\hat{\sigma} )=\delta (\hat{\sigma})$, which leads to the self-consistent equation of the cage size:  
\begin{equation}
  \frac{1}{\hat{A} (\hat{\sigma},\overline{\phi} )}= \overline{\phi} e^{\hat{\sigma}/2} g \left( \frac{\hat{A} (\hat{\sigma},\overline{\phi} )+\hat{A} (0,\overline{\phi})}{2}\right).  
  \label{detA_NM}
\end{equation}
For $\hat{\sigma} = 0$, this equation is reduced to 
\begin{equation}
  \frac{1}{\hat{A} (0,\overline{\phi})}= \overline{\phi} g \left(\hat{A}(0,\overline{\phi}) \right), 
  \label{detA_M}
\end{equation} 
which is precisely the same as the cage size equation for monodisperse hard spheres~\cite{ParisiBook}.

\begin{figure}[t]
  \centering
     \includegraphics[width=0.95\columnwidth]{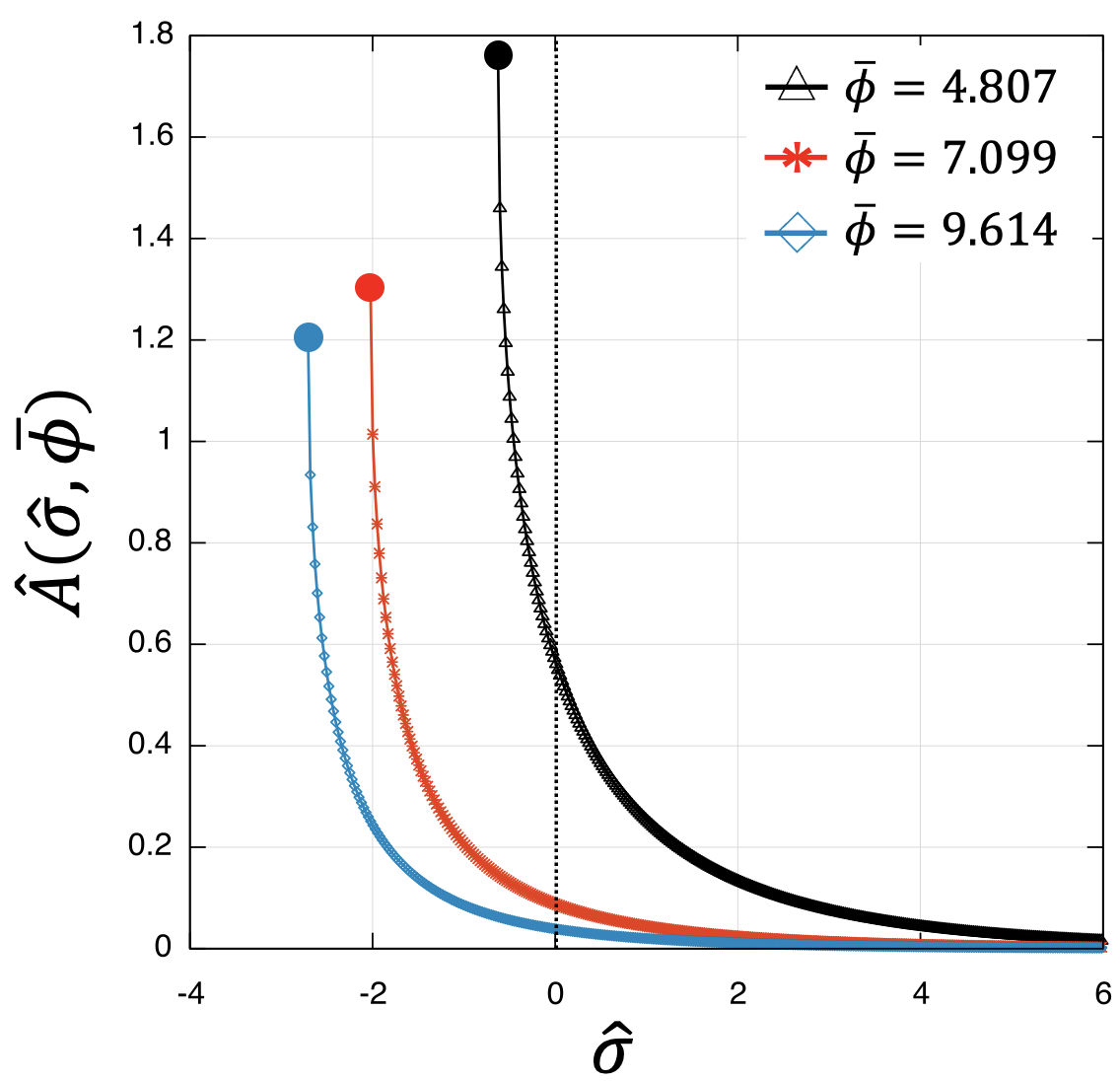}
     \caption{
The cage size with respect to the particle size for the nearly monodisperse system at three densities: the dynamic glass transition density $\overline{\phi} = 4.807 = \overline{\phi}_{\rm d}$ (triangles), $7.099$ (asterisks), and $9.614$ (squares). 
The filled circles indicate the critical particle size $\hat{\sigma}^{\ast}$. 
     }
     \label{nmA}
\end{figure}

\subsection{Numerical Results} 
 
First, we numerically solved Eqs.~(\ref{detA_NM}) and (\ref{detA_M}) via an iterative algorithm. 
We first solved Eq.~(\ref{detA_M}) for the packing fraction $\overline{\phi}$ to obtain $\hat{A}(0,\overline{\phi})$. 
We then substituted it into Eq.~(\ref{detA_NM}) and solved to obtain $\hat{A}(\hat{\sigma},\overline{\phi})$. 
We repeated this calculation at various $\overline{\phi}$. 
We find that the cage size becomes finite only when $\hat{A}(0,\overline{\phi})$ is finite.
As a result, the dynamic glass transition density of the NM system is the same as that of the monodisperse system, $\overline{\phi}_{\rm d}=4.807$.

In Fig.~\ref{nmA}, we plot the cage size $\hat{A}(\hat{\sigma},\overline{\phi})$ with respect to the particle size $\hat{\sigma}$ for several packing fractions. 
The triangles represent the cage size at the dynamic glass transition density $\overline{\phi} = \overline{\phi}_{\rm d}$. 
The cage size is finite only in $\hat{\sigma} \geq -0.627$. 
This means that larger particles with $\hat{\sigma} \geq -0.627$ are trapped in the cage and behave as a glass, while smaller particles with $\hat{\sigma} < -0.627$ behave as a liquid. 
This dynamic decoupling is mentioned in the Introduction.
We define the critical particle size $\hat{\sigma}^{\ast}$, above which the cage size is finite. 
At this density $\overline{\phi} = \overline{\phi}_{\rm d}$, the critical particle size is $\hat{\sigma}^{\ast}=-0.627$. 

The asterisks and squares in Fig.~\ref{nmA} represent the cage sizes at two higher densities.
Increasing the density decreases the cage size and the critical particle size $\hat{\sigma}^{\ast}$, thus causing the vitrification of smaller particles.

\begin{figure*}[t]
  \centering
     \includegraphics[width=0.95\linewidth]{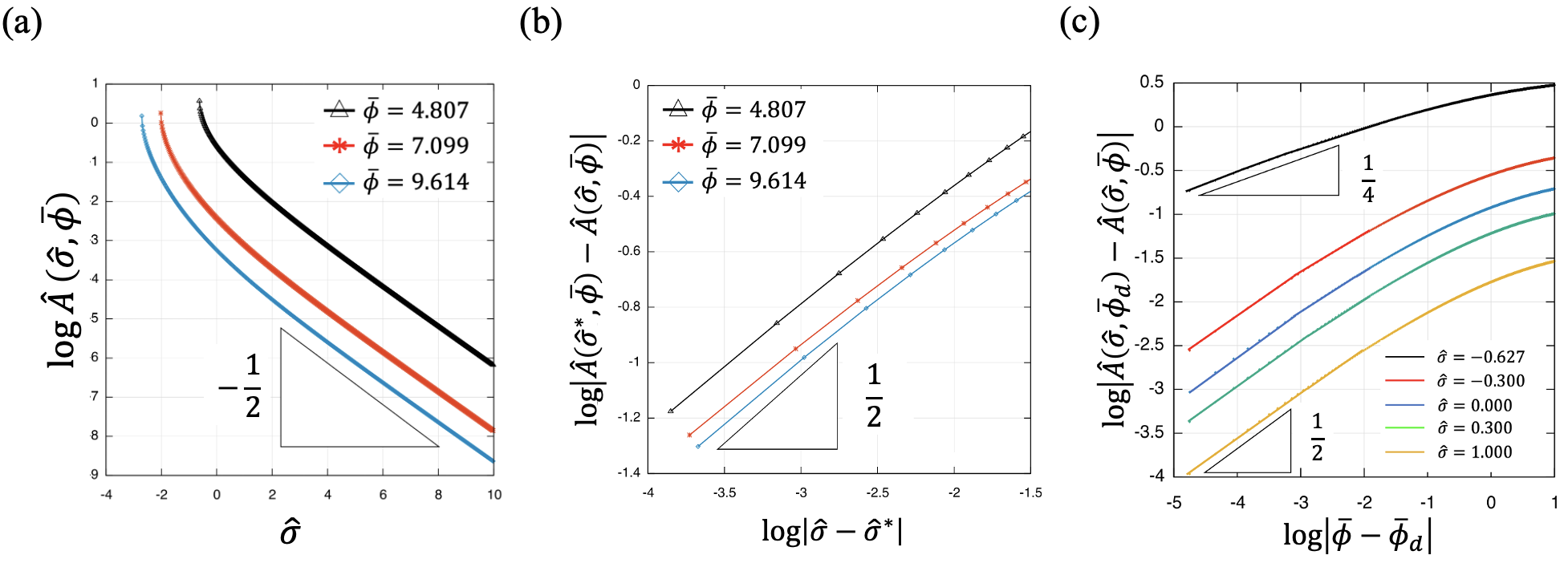}
     \caption{
Asymptotic laws of the cage size in the nearly-monodisperse system. 
(a)~The cage size $\hat{A}(\hat{\sigma},\overline{\phi})$ with respect to the particle size $\hat{\sigma}$ in a semilog plot. 
The slope $-1/2$ indicates the exponential law given in Eq.~(\ref{NM_A1/2}).
(b)~$\log |\hat{A}(\hat{\sigma}^{\ast},\overline{\phi}) - \hat{A}(\hat{\sigma},\overline{\phi})|$ with respect to $\log |\hat{\sigma} - \hat{\sigma}^{\ast}|$ in the vicinity of the critical size $\hat{\sigma} \simeq \hat{\sigma}^{\ast}$. 
The slope $1/2$ indicates the square-root singularity Eq.~(\ref{NM_cri_size}).
(c)~$\log |\hat{A}(\hat{\sigma},\overline{\phi}_{\rm d}) - \hat{A}(\hat{\sigma},\overline{\phi})|$ with respect to $\log |\overline{\phi} - \overline{\phi}_{\rm d}|$ in the vicinity of the dynamic glass transition $\overline{\phi} \simeq \overline{\phi}_{\rm d}$. 
The data are for five representative $\hat{\sigma}$ including $\hat{\sigma}^{\ast} = -0.627$ at $\overline{\phi} = \overline{\phi}_{\rm d}$. 
The slopes $1/4$ and $1/2$ indicate the critical laws shown in Eq.~(\ref{NM_cri_den}).
     }
     \label{nmAS1/2}
\end{figure*} 

\subsection{Asymptotic analysis} 

In this subsection, we analyze the self-consistent equations Eqs.~(\ref{detA_NM}) and (\ref{detA_M}) semianalytically. 
In particular, we derive several asymptotic laws of cage size in the NM system. 
 
\subsubsection{Large particles: $\hat{\sigma} \gg 1$} 

First, we focus on large particles with $\hat{\sigma} \gg 1$. 
Fig.~\ref{nmA} shows that the cage size rapidly decreases with $\hat{\sigma}$ in this regime. 
We derive an asymptotic law of this behavior. 

In Eq.~(\ref{detA_NM}), the argument of the function $g(x)$ is the average of the two cage sizes: $(\hat{A}(\hat{\sigma},\overline{\phi}) + \hat{A}(0,\overline{\phi}))/2$. 
We now assume $\hat{A}(\hat{\sigma},\overline{\phi}) \ll \hat{A}(0,\overline{\phi})$, so that the average can be approximated by $\hat{A}(0,\overline{\phi})/2$. 
Under this condition, the function $g$ is independent of $\hat{\sigma}$, and Eq.~(\ref{detA_NM}) is reduced to 
\begin{equation}
\hat{A}(\hat{\sigma},\overline{\phi}) \propto  e^{- \hat{\sigma}/2}. 
\label{NM_A1/2}
\end{equation}
This means that the cage size decreases exponentially with the particle size. 
Because this verifies the assumption $\hat{A}(\hat{\sigma},\overline{\phi}) \ll \hat{A}(0,\overline{\phi})$ at $\hat{\sigma} \gg 1$, the exponential law Eq.~(\ref{NM_A1/2}) is valid at $\hat{\sigma} \gg 1$. 
 
To test the exponential law of Eq.~(\ref{NM_A1/2}) precisely, we plotted $\log{\hat{A} (\hat{\sigma},\overline{\phi})}$ with respect to $\hat{\sigma}$ in Fig.~\ref{nmAS1/2}a in a semilog plot. 
We find a linear relationship with the slope $-1/2$ for all the densities, confirming the exponential law.

\subsubsection{In the vicinity of the critical size: $\hat{\sigma} \simeq \hat{\sigma}^*$} 
 
Next, we focus on the particles with size close to the critical size, $\hat{\sigma} \simeq \hat{\sigma}^{\ast}$.
Fig.~\ref{nmA} shows that the cage size drastically decreases with a slight increase in $\hat{\sigma}$.
We analyze this behavior.

\begin{figure}[t]
  \centering
     \includegraphics[width=0.95\columnwidth]{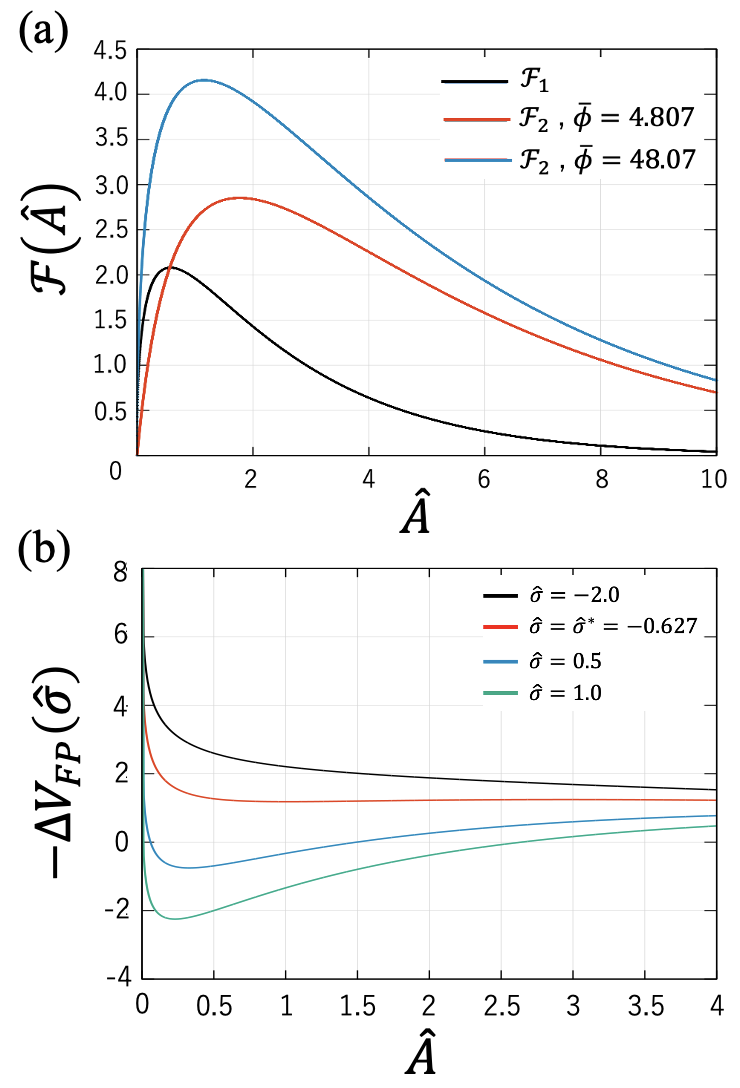}
     \caption{
(a)~The auxiliary functions $\mathcal{F}_1 (\hat{A})$ and $\mathcal{F}_2 (\hat{A};\overline{\phi})$. 
(b)~$\Delta V_{\rm FP}(\hat{\sigma})$ which expresses the contribution to the FP potential from each particle size $\hat{\sigma}$. 
The data are for four representative $\hat{\sigma}$ at $\overline{\phi} = \overline{\phi}_{\rm d}$. 
The stationary point of $\Delta V_{\rm FP}(\hat{\sigma})$ at finite $\hat{A}$ corresponds to the solution of Eq.~(\ref{detA_NM}). 
     }
     \label{F1F2}
\end{figure}

Following the anlaysis for the monodisperse system~\cite{ParisiReview}, we first transform Eq.~(\ref{detA_M}) into the following form: 
\begin{equation}
  \frac{1}{\overline{\phi}} = \mathcal{F}_{1} \left(\hat{A}(0,\overline{\phi})\right) \label{F1}
\end{equation}
where we introduce an auxiliary function $\mathcal{F}_1 (x) = x g(x)$. 
According to Eq.~(\ref{F1}), the intersection of $y= \mathcal{F}_1 (x)$ and the horizontal line $y=1/\overline{\phi}$ determines $\hat{A}(0,\overline{\phi})$. 
Because the function $\mathcal{F}_1 (x)$ has only one maximum (see Fig.~\ref{F1F2}a), the dynamic glass transition density is determined by $1/\overline{\phi}_{\rm d} = \max_x \mathcal{F}_1 (x)$. 

We consider a similar transformation of Eq.~(\ref{detA_NM}). 
When we fix the packing fraction at $\overline{\phi} \geq \overline{\phi}_{\rm d}$, we can treat $\hat{A}(0,\overline{\phi})$ as a finite constant. 
Therefore, Eq.~(\ref{detA_NM}) can be transformed into 
\begin{equation}
\frac{e^{-\hat{\sigma}/2}}{\overline{\phi}} = \mathcal{F}_{2} \left( \hat{A}(\hat{\sigma},\overline{\phi}); \overline{\phi}\right) \label{F2}
\end{equation}
where we introduce an auxiliary function $\mathcal{F}_2 (x; \overline{\phi}) = x g\left( (x+\hat{A}(0,\overline{\phi}))/2 \right)$. 
As shown in Fig.~\ref{F1F2}a, this auxiliary function $\mathcal{F}_2 (x; \overline{\phi})$ behaves quite similarly to $\mathcal{F}_1 (x)$. 
The intersection of $y=\mathcal{F}_{2}(x;\overline{\phi})$ and the horizontal line $y=e^{-\hat{\sigma}/2}/\overline{\phi}$ determines the cage size $\hat{A}(\hat{\sigma},\overline{\phi})$ for any $\hat{\sigma}$. 
Accordingly, the critical particle size $\hat{\sigma}^{\ast}$ is determined by 
\begin{equation}
e^{-\hat{\sigma}^{\ast}/2} = \overline{\phi} \max_x \mathcal{F}_2 (x;\overline{\phi}). 
\end{equation}

Now, we focus on the vicinity of the critical size.
By introducing a simplified notation $\hat{A}^{\ast}=\hat{A}\left(\hat{\sigma}^{\ast},\overline{\phi}\right)$, we expand the particle size as $\hat{\sigma}=\hat{\sigma}^{\ast}+\Delta \hat{\sigma}$ and the cage size as $\hat{A}\left(\hat{\sigma},\overline{\phi}\right) = \hat{A}^{\ast} - \Delta \hat{A}$. 
Substituting these into Eq.~(\ref{F2}) and expanding both sides, we obtain 
\begin{eqnarray}
- \frac{e^{-\frac{\hat{\sigma}^{\ast}}{2}}}{2 \overline{\phi}} \Delta \hat{\sigma} = - \mathcal{F}_{2}'(\hat{A}^{\ast};\overline{\phi}) \Delta \hat{A} + \frac{1}{2} \mathcal{F}_{2}''(\hat{A}^{\ast};\overline{\phi}) \Delta \hat{A}^2
\end{eqnarray}
in the lowest orders of $\Delta \hat{\sigma}$ and $\Delta \hat{A}$. 
The prime symbol indicates the derivative of $\mathcal{F}_{2}(x;\overline{\phi})$ with respect to $x$. 
Because $\hat{A}^{\ast}$ corresponds to the maximum of $\mathcal{F}_{2} (x;\overline{\phi})$, the first derivative vanishes $\mathcal{F}_{2}'(\hat{A}^{\ast};\overline{\phi})=0$ and the second derivative is negative $\mathcal{F}_{2}''(\hat{A}^{\ast};\overline{\phi}) < 0$.
Therefore, we obtain $\Delta \hat{\sigma} \propto \Delta \hat{A}^2$, or equivalently,  
\begin{equation}
\hat{A}\left(\hat{\sigma}^{\ast},\overline{\phi}\right) - \hat{A}\left(\hat{\sigma},\overline{\phi}\right) \propto  (\hat{\sigma}-\hat{\sigma}^{\ast})^{1/2}. \label{NM_cri_size}
\end{equation}
Because this argument is independent of the density, Eq.~(\ref{NM_cri_size}) should be valid in any $\overline{\phi} \geq \overline{\phi}_{\rm d}$. 

The square-root singularity shown in Eq.~(\ref{NM_cri_size}) indicates that $\hat{\sigma}^{\ast}$-sized particles are on the verge of spinodal instability. 
To discuss this point directly, we investigated the FP potential. 
By substituting Eq.~(\ref{del_dist}) into Eq.~(\ref{FP}), we obtain the FP potential of the NM system. 
In the leading order of $\epsilon$, this FP potential can be decomposed into the contribution from each particle size. 
With the omission of unimportant constants, the contribution from $\hat{\sigma}$ is 
\begin{align} 
& \Delta V_{\rm FP}(\hat{\sigma}) = \log{2\pi \hat{A}(\hat{\sigma})}  \nonumber \\
& + 2 \overline{\phi}_{0} e^{\hat{\sigma}/2} \sqrt{4 \hat{A}_{\rm ave}} e^{-\hat{A}_{\rm ave}} \int_{-\infty}^{\infty} dz e^{\sqrt{4\hat{A}_{\rm ave}}z} \varTheta (z) \log{\varTheta(z)}
\label{FPdec}
\end{align}
where $\hat{A}_{\rm ave} = \frac{\hat{A}(0) + \hat{A}(\hat{\sigma})}{2}$~\footnote{ 
Note that we can reproduce the self-consistent equation Eq.~(\ref{detA_NM}) by imposing the stationary point condition on Eq.~(\ref{FPdec}).}.
In Fig.~\ref{F1F2}b, we plot $\Delta V_{\rm FP}(\hat{\sigma})$ for four representative particle sizes at the density $\overline{\phi} = \overline{\phi}_{\rm d}$. 
$\Delta V_{\rm FP}(\hat{\sigma})$ has a stationary point for larger particles $\hat{\sigma} > \hat{\sigma}^{\ast}$ but not for smaller particles $\hat{\sigma} < \hat{\sigma}^{\ast}$. 
At the critical size $\hat{\sigma} = \hat{\sigma}^{\ast}$, a stationary point emerges at which the FP potential becomes flat. 
This clearly shows that $\hat{\sigma}^{\ast}$-sized particles are on the verge of spinodal instability. 
For monodisperse systems, the same instability is known to emerge at the dynamic glass transition density $\overline{\phi} = \overline{\phi}_{\rm d}$. 
By contrast, for the NM system, this instability is ubiquitous at $\overline{\phi} \geq \overline{\phi}_{\rm d}$ since there are always $\hat{\sigma}^{\ast}$-sized particles in the system. 
 
To test the square-root law of Eq.~(\ref{NM_cri_size}), we plotted $\log{|\hat{A}\left(\hat{\sigma},\overline{\phi}\right)-\hat{A}\left(\hat{\sigma}^{\ast},\overline{\phi}\right)|}$ with respect to $\log{|\hat{\sigma}-\hat{\sigma}^{\ast}|}$ in Fig.~\ref{nmAS1/2}b. 
The slope is $1/2$ for all the densities, confirming the square-root law.

 \subsubsection{Density dependence}

Finally, we analyze the density dependence of the cage size in the vicinity of the dynamic glass transition, $\overline{\phi} \simeq \overline{\phi}_{\rm d}$.
To this end, we expand the density as $\overline{\phi} = \overline{\phi}_{\rm d} + \Delta \overline{\phi}$.
Furthermore, we expand the cage size as $\hat{A}(\hat{\sigma},\overline{\phi}) = \hat{A}_{\rm d} - \Delta \hat{A}$ and $\hat{A}(0,\overline{\phi}_{\rm d}) = \hat{A}_{\rm 0d} - \Delta \hat{A}_0$, introducing simplified notations $\hat{A}_{\rm d}=\hat{A}(\hat{\sigma},\overline{\phi}_{\rm d})$ and $\hat{A}_{\rm 0d}=\hat{A}(0,\overline{\phi}_{\rm d})$. 
By substituting these expressions into Eqs.~(\ref{F1}) and (\ref{F2}) and expanding them, we can obtain the critical law in the vicinity of the dynamic glass transition. 

We first focus on $\hat{\sigma}=0$, where the argument is precisely the same as in the monodisperse case.
In this case, we can expand Eq.~(\ref{F1}) as
\begin{eqnarray}
- \frac{1}{\overline{\phi}_{\rm d}^{2}}\Delta \overline{\phi} = \frac{1}{2} \mathcal{F}_{1}''(\hat{A}_{\rm 0d}) \Delta \hat{A}_0^2
\end{eqnarray}
in the leading order of $\Delta \overline{\phi}$ and $\Delta \hat{A}_0$. 
Therefore, we obtain $\Delta \hat{A}_0 \propto \Delta \overline{\phi}^{1/2}$. 

We then focus on $\hat{\sigma} \neq 0$. 
In this case, we need to calculate the variation in $\mathcal{F}_2 (x; \overline{\phi})$ in both $x$ and $\overline{\phi}$. 
This can be done by observing 
\begin{align} \label{intermediate}
\mathcal{F}_2 \left(\hat{A}(\hat{\sigma},\overline{\phi}); \overline{\phi} \right) & = (\hat{A}_{\rm d} - \Delta \hat{A}) \nonumber \\
& \times g \left( \frac{\hat{A}_{\rm d} + \hat{A}_{\rm 0d}}{2} - \frac{\Delta \hat{A} +\Delta \hat{A}_0}{2} \right). 
\end{align}
Expanding Eq.~(\ref{intermediate}) in $\Delta \hat{A}$ and $\Delta \hat{A}_0$ and substituting it into Eq.~(\ref{F2}), we obtain
\begin{align}
- \frac{e^{-\frac{\hat{\sigma}}{2}}}{\overline{\phi}_{\rm d}^{2}}\Delta \overline{\phi}  &= - \mathcal{F}_{2}'(\hat{A}_{\rm d}; \overline{\phi}_{\rm d}) \Delta \hat{A} +  \frac{1}{2}\mathcal{F}_{2}''(\hat{A}_{\rm d};\overline{\phi}_{\rm d}) \Delta \hat{A}^2 \nonumber \\
& \quad - \hat{A}_{\rm d} g'\left(\frac{\hat{A}_{\rm d} + \hat{A}_{\rm 0d}}{2}\right) \frac{\Delta \hat{A}_0}{2}
\label{variation}
\end{align}
for the relevant terms. 
In this equation, the lowest order of $\Delta \overline{\phi}$ is not found on the left-hand side, but instead is the third term on the right-hand side because $\Delta \hat{A}_0 \propto \Delta \overline{\phi}^{1/2}$. 
The first term and the third term on the right-hand side should be balanced, which results in $\Delta \hat{A} \propto \Delta \hat{A}_0 \propto \Delta \overline{\phi}^{1/2}$.
However, if $\hat{\sigma} = \hat{\sigma}^{\ast}$, the first term vanishes as discussed in the previous subsection.
Thus, in this case, the second and the third terms on the right-hand side should be balanced, which results in $\Delta \hat{A}^2 \propto \Delta \hat{A}_0 \propto \Delta \overline{\phi}^{1/2}$.
Summarizing these results, we obtain the following critical law for the density dependence of the cage size:  
\begin{equation}
\hat{A}\left(\hat{\sigma},\overline{\phi}_{\rm d}\right) - \hat{A}\left(\hat{\sigma},\overline{\phi}\right) \propto  
\begin{cases}
(\overline{\phi}-\overline{\phi}_{\rm d})^{1/2} & \text{if $\hat{\sigma} > \hat{\sigma}^{\ast}$,} \\
(\overline{\phi}-\overline{\phi}_{\rm d})^{1/4} & \text{if $\hat{\sigma}=\hat{\sigma}^{\ast}$.} \label{NM_cri_den}
\end{cases}
\end{equation}
in the vicinity of the dynamic glass transition. 
Interestingly, the density dependence is modified for $\hat{\sigma}^{\ast}$-sized particles, which is due to the coexistence of two criticalities associated with the dynamic glass transition density and the critical particle size. 

To test the critical laws Eq.~(\ref{NM_cri_den}), we plotted $\log{|\hat{A}\left(\hat{\sigma},\overline{\phi}\right)-\hat{A}\left(\hat{\sigma},\overline{\phi}_{\rm d}\right)|}$ with respect to $\log{|\overline{\phi}-\overline{\phi}_{\rm d}|}$ in Fig.~\ref{nmAS1/2}c. 
The slope is $1/2$ for $\hat{\sigma} \neq \hat{\sigma}^{\ast}$, while the slope is $1/4$ for $\hat{\sigma}=\hat{\sigma}^{\ast}$, which confirms the critical law Eq.~(\ref{NM_cri_den}). 

\subsubsection{Summary}

We summarize the results of the asymptotic analysis for the NM system.
At a fixed density, the cage size decreases exponentially with the particle size at $\hat{\sigma} \gg 1$ (Eq.~(\ref{NM_A1/2})) and decreases in the power law with the exponent $1/2$ at $\hat{\sigma} \simeq \hat{\sigma}^{\ast}$ (Eq.~(\ref{NM_cri_size})). 
This square-root singularity originates from spinodal instability of $\hat{\sigma}^{\ast}$-sized particles, as depicted in Fig.~\ref{F1F2}b. 
When the density is varied, the cage size decreases with the density in the power-law with the exponent $1/4$ for $\hat{\sigma}^{\ast}$-sized particles and $1/2$ for other particles (Eq.~(\ref{NM_cri_den})).

\section{Exponential distribution system}

The NM system studied in the previous section is characterized by an extremely narrow particle size distribution. 
In this section, we focus on a more polydisperse system. 
Specifically, we study exponential distribution (ED) systems in which the particle size follows an exponential distribution: 
\begin{equation}
\hat{\varphi}\left(\hat{\sigma}\right) = 
\begin{cases}
\mathcal{N} e^{-\gamma \hat{\sigma}} & \text{if $0 \leq \hat{\sigma} \leq \hat{\sigma}_{\rm M}$,} \\
0   & \text{otherwise.} 
\end{cases}
\label{exp_dist}
\end{equation}
Here, $\hat{\sigma}_{\rm M}$ represents the upper cutoff of the particle size, and $\gamma$ is a parameter related to the width of the distribution. 
At smaller $\gamma$, the distribution becomes broader, and the system becomes more polydisperse. 
We refer to $\gamma$ as the width parameter. 
The normalization constant $\mathcal{N}$ satisfies $1-e^{-\gamma \hat{\sigma}_{\rm M}} =\gamma/\mathcal{N}$. 
In the following, we mainly study the case of $\hat{\sigma}_{\rm M} = 5$ and discuss $\hat{\sigma}_{\rm M}$-dependence later.

\subsection{Interpretation in the finite dimensions}

First, we argue that the ED system is an infinite-dimensional counterpart of the power-law distribution systems in finite dimensions.
To this end, we consider a continuously polydisperse system in the $d$-dimension, where the particle size follows the distribution:
\begin{equation}
  \varphi\left(\sigma\right) \varpropto  \sigma^{-a}. \label{power_dist}
\end{equation}
This distribution appears frequently in nature. 
For example, in foams~\cite{shimamoto}, grains in rocks, and sediments formed by an impact fracture~\cite{SammisRock,impactOdder,Ryugu}, the particle size distribution is often described by Eq.~(\ref{power_dist}).  
Additionally, in recent simulation studies on the glass transition, Eq.~(\ref{power_dist}) has been frequently adopted to accelerate numerical simulations~\cite{Ninarello2017}.

Substituting the particle size scaling Eq.~(\ref{sizescaling}) into Eq.~(\ref{power_dist}) and taking $d \to \infty$ while keeping $a/d$ constant, we obtain 
\begin{equation}
  \sigma^{-a} \varpropto \left(1+\frac{\hat{\sigma}}{d}\right)^{-a} 
  \to \exp \left[ - (a/d) \hat{\sigma} \right]. 
\end{equation}
Therefore, the exponential distribution Eq.~(\ref{exp_dist}) is the infinite-dimensional counterpart of the power-law distribution Eq.~(\ref{power_dist}), where the width parameter $\gamma$ is given by $a = \gamma d$.

\begin{figure}[t]
  \centering
     \includegraphics[width=0.95\columnwidth]{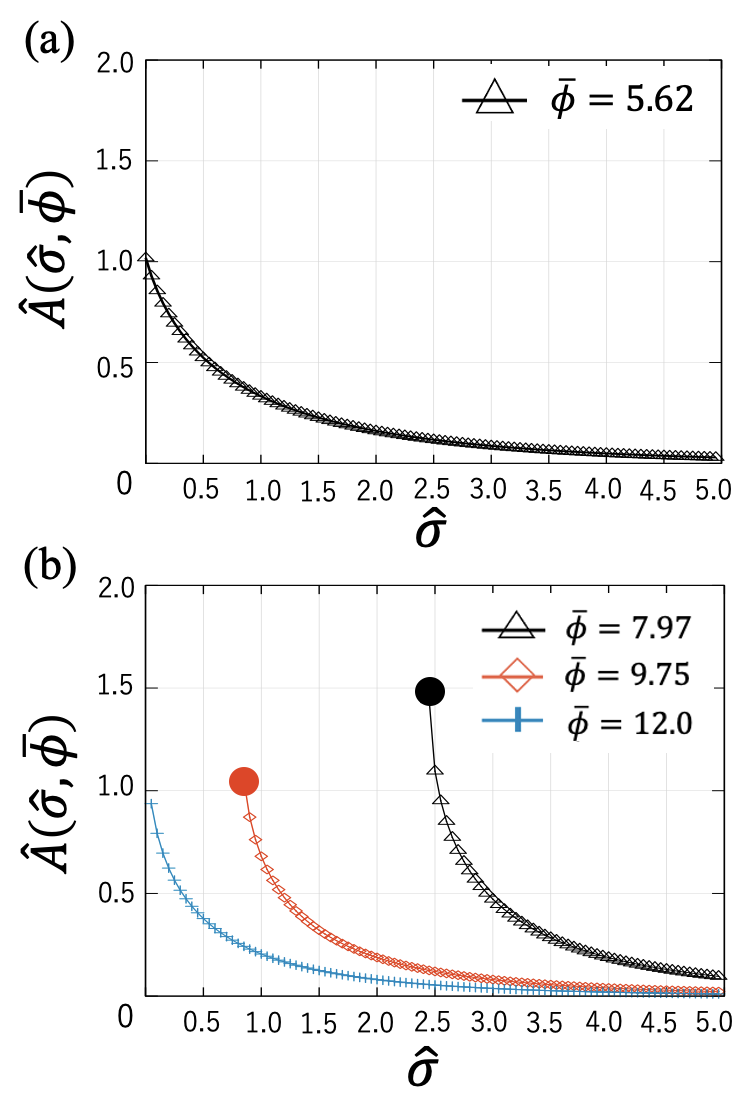}
     \caption{
The cage size with respect to the particle size for the exponential distribution system. 
(a)~For $\gamma = 2$ at the dynamic glass transition density $\overline{\phi}=\overline{\phi}_{\rm d} = 5.62$. 
The cage size is finite for all the particles. 
(a)~For $\gamma = 0.5$ at three densities: the dynamic glass transition density $\overline{\phi}=\overline{\phi}_{\rm d} = 7.97$, $\overline{\phi}=9.75$, and the complete dynamic glass transition density $\overline{\phi}=\overline{\phi}_{\rm d,all} = 12.0$. 
The filled circles indicate the critical particle size $\hat{\sigma}^{\ast}$. 
    }
     \label{exp_cutoff_AS}
\end{figure}

\subsection{Numerical Results}

Substituting the exponential distribution Eq.~(\ref{exp_dist}) into the self-consistent equation Eq.~(\ref{determiningA}), we obtain the integral equation for the cage size $\hat{A}(\hat{\sigma},\overline{\phi})$. 
We solved this equation numerically via an iterative method.
We first show the results for two representative cases: $\gamma=2$ and $\gamma=0.5$. 

\subsubsection{Narrow distribution: $\gamma=2$}

We solved the self-consistent equation for $\gamma=2$ at various densities. 
We found that the cage size becomes finite at $\overline{\phi} \geq 5.62$; therefore, the dynamic glass transition density is $\overline{\phi}_{\rm d}=5.62$. 
Fig.~\ref{exp_cutoff_AS}a shows the cage size at the dynamic glass transition density. 
The cage size is finite in the entire domain of the particle size: $0 \leq \hat{\sigma} \leq \hat{\sigma}_{\rm M} = 5$. 
This means that all the particles in the system simultaneously vitrified.

\subsubsection{Broad distribution: $\gamma=0.5$}

Next, we solved the self-consistent equation for $\gamma=0.5$ at various densities. 
We found that the cage size becomes finite only at $\overline{\phi} \geq \overline{\phi}_{\rm d} = 7.97$. 
Fig.~\ref{exp_cutoff_AS}b shows the cage size at several densities. 
At the dynamic glass transition density, the cage size is finite for large particles, but not for small particles. 
Therefore, dynamic decoupling occurs.
Following the analysis for the NM system, we define the critical particle size $\hat{\sigma}^{\ast}$ above which the cage size is finite.
We emphasize that $\hat{\sigma}^{\ast}$ differs from the lower cutoff $\hat{\sigma} = 0$ and the upper cutoff $\hat{\sigma} = \hat{\sigma}_{\rm M}$ of the particle size distribution. 
In other words, when the system vitrifies, the system spontaneously defines the ``larger'' particles via the self-consistent equation. 

At higher densities, the cage size decreases, and the critical particle size $\hat{\sigma}^{\ast}$ also decreases.
At $\overline{\phi} = 12.0$, the critical particle size vanishes $\hat{\sigma}^{\ast} = 0$, meaning that all the particles vitrify at this density. 
We define the complete dynamic glass transition density $\overline{\phi}_{\rm d,all}$ as the density above which the cage size is finite in the entire domain of the particle size. 
The complete dynamic glass transition density is $\overline{\phi}_{\rm d,all} = 12.0$. 

\begin{figure}[t]
  \centering
     \includegraphics[width=0.95\columnwidth]{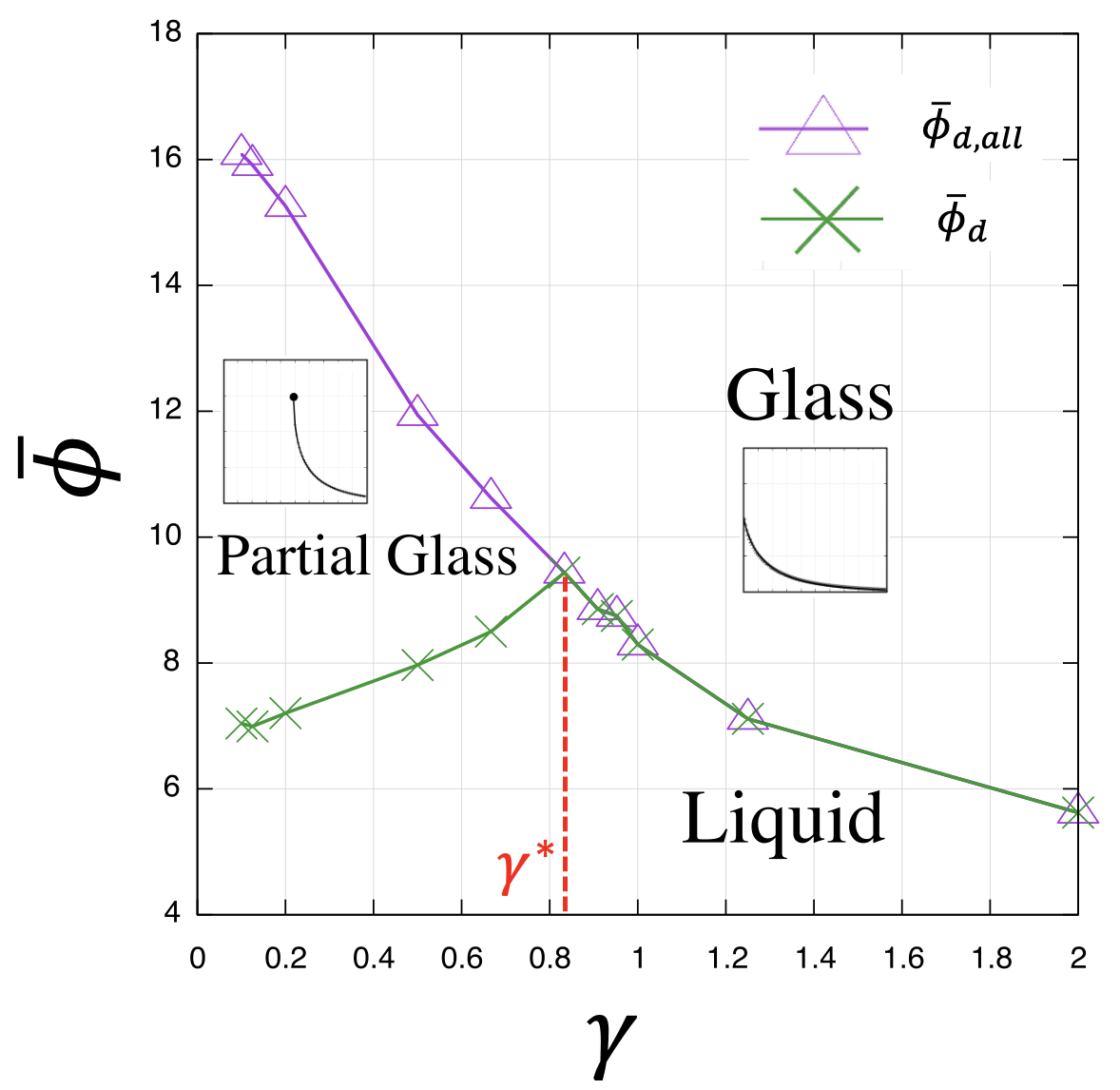}
     \caption{
Phase diagram of the exponential distribution system in the packing fraction $\overline{\phi}$ and the width parameter $\gamma$. 
The green crosses are the dynamic glass transition density $\overline{\phi}_{\rm d}$ below which none of the particles have finite cage size. 
The purple triangles are the complete dynamic glass transition density $\overline{\phi}_{\rm d,all}$ above which all the particles have a finite cage size. 
All the particles vitrify in the ``Glass'' phase, while only larger particles vitrify in the ``Partial Glass'' phase.
The inset shows the typical behavior of the cage size in each phase. 
$\gamma^{\ast}$ indicates the critical width parameter at which two glass transition lines merge. 
     }
     \label{exp_cutoff_phase}
\end{figure}

\begin{figure*}[t]
  \centering
     \includegraphics[width=0.95\linewidth]{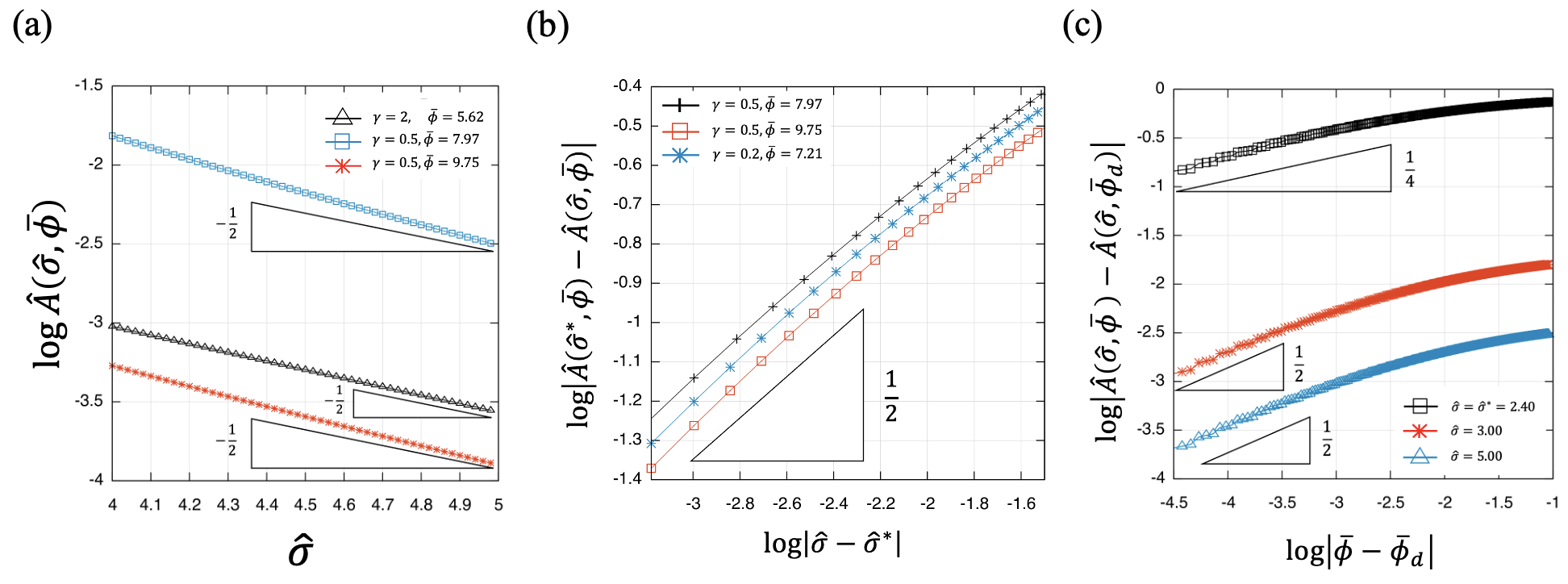}
     \caption{
This is the same as Fig.~\ref{nmAS1/2}, but for the exponential distribution system. 
The data in the panel (c) are for $\gamma = 0.5$. 
     }
     \label{FIG_EXP_cri}
\end{figure*}

\subsubsection{Phase diagram}

We repeated the cage size calculations at various $\gamma$ and $\overline{\phi}$.
We summarize the results in Fig.~\ref{exp_cutoff_phase} as a phase diagram.
The green crosses and the purple triangles show the dynamic glass transition density $\overline{\phi}_{\rm d}$ and the complete dynamic glass transition density $\overline{\phi}_{\rm d,all}$, respectively. 
For any $\gamma$, the system is in a liquid state at low densities, while the cage size becomes finite, and the system vitrifies at high densities.
When $\gamma$ is large (cf. $\gamma = 2$), the dynamic glass transition density $\overline{\phi}_{\rm d}$ coincides with the complete dynamic glass transition density $\overline{\phi}_{\rm d,all}$. 
Namely, the cage sizes for all the particles become finite simultaneously. 
On the other hand, when $\gamma$ is small (cf. $\gamma= 0.5$), we find $\overline{\phi}_{\rm d,all} > \overline{\phi}_{\rm d}$.
Namely, larger particles vitrify first, and smaller particles vitrify only at higher densities.
We refer to the state at $\overline{\phi} \geq \overline{\phi}_{\rm d,all}$ as ``glass'', where all the particles vitrify, and the state at $\overline{\phi}_{\rm d,all} > \overline{\phi} \geq \overline{\phi}_{\rm d}$ as ``partial glass'', where only larger particles vitrify. 
The two glass transition lines merge at $\gamma = \gamma^{\ast} \simeq 0.82$, which we call the critical width parameter. 
The partial glass phase appears only when the size distribution is sufficiently broad, $\gamma < \gamma^{\ast}$.

\subsection{Asymptotic Analysis}

In Sec.~III, we derived several asymptotic laws of the cage size in the NM system. 
Here, we discuss the validity of these laws in the ED system. 
The semianalytical approach for the NM system cannot be applied to the ED system because $\hat{A}(\hat{\sigma})$ is coupled with all other $\hat{A}(\hat{\sigma}')$ via the self-consistent equation Eq.~(\ref{determiningA}). 
However, we show that the cage size of the ED system still follows the same asymptotic laws, suggesting that the understanding of the NM system is effective even for the ED system. 

First, we focus on the cage size of large particles. 
In the NM system, the cage size decreases exponentially with increasing particle size, as Eq.~(\ref{NM_A1/2}).
To verify this exponential law in the ED system, we plotted the logarithm of the cage size with respect to the particle size in Fig.~\ref{FIG_EXP_cri}a. 
For the state ($\gamma = 2$, $\overline{\phi} = 5.62$) in the Glass phase, the slope $-1/2$ works well, suggesting the validity of the exponential law. 
However, the description by the exponential law becomes slightly worse for ($\gamma = 0.5$, $\overline{\phi} = 9.75$) and worse for ($\gamma = 0.5$, $\overline{\phi}=7.97$) in the partial glass phase. 
This deviation occurs because the focused particle size $\hat{\sigma} \lesssim 5$ is not far from the critical particle size $\hat{\sigma}^{\ast}$ in these states, and the results suffer from the square-root singularity associated with $\hat{\sigma}^{\ast}$. 
Therefore, we conclude that the exponential law Eq.~(\ref{NM_A1/2}) is valid for $\hat{\sigma} \gg 1$ as long as $\hat{\sigma} \gg \hat{\sigma}^{\ast}$. 

Next, we focus on the vicinity of the critical particle size, $\hat{\sigma} \simeq \hat{\sigma}^{\ast}$.
In the NM system, the cage size exhibits the square-root singularity shown in Eq.~(\ref{NM_cri_size}), which originates from the spinodal instability at $\hat{\sigma} = \hat{\sigma}^{\ast}$.
To verify the square-root singularity in the ED system, we plotted $\log{|\hat{A}\left(\hat{\sigma},\overline{\phi}\right)-\hat{A}\left(\hat{\sigma}^{\ast},\overline{\phi}\right)|}$ with respect to $\log{|\hat{\sigma}-\hat{\sigma}^{\ast}|}$ in Fig.~\ref{FIG_EXP_cri}b. 
The data are for several state points in the partial glass phase, where the critical particle size is finite.
Clearly, all the data are consistent with the slope $1/2$, and thus, the square-root singularity Eq.~(\ref{NM_cri_size}) is valid even in the ED system. 

Finally, we focus on the density dependence of the cage size. 
In the NM system, the cage size follows the critical laws Eq.~(\ref{NM_cri_den}): the power-laws with exponents $1/2$ for $\hat{\sigma} \neq \hat{\sigma}^{\ast}$ and $1/4$ for $\hat{\sigma} = \hat{\sigma}^{\ast}$. 
To verify this in the ED system, we plotted $\log{|\hat{A}\left(\hat{\sigma},\overline{\phi}\right)-\hat{A}\left(\hat{\sigma},\overline{\phi}_{\rm d}\right)|}$ with respect to $\log{|\overline{\phi}-\overline{\phi}_{\rm d}|}$ in Fig.~\ref{FIG_EXP_cri}c. 
Clearly, the slope is $1/2$ for $\hat{\sigma} \neq \hat{\sigma}^{\ast}$ and $1/4$ for $\hat{\sigma}=\hat{\sigma}^{\ast}$. 
Therefore, the critical laws of Eq.~(\ref{NM_cri_den}) are valid even in the ED system. 

In summary, all the critical laws derived for the NM system are valid in the ED system. 
The validity of the square-root singularity Eq.~(\ref{NM_cri_size}) suggests that $\hat{\sigma}^{\ast}$-sized particles are on the verge of spinodal instability (see Fig.~\ref{F1F2}b), even in the ED system.
Because there are always $\hat{\sigma}^{\ast}$-sized particles in the system in the partial glass phase, this instability is ubiquitous in this phase, which is in sharp contrast to the monodisperse system.

\subsection{Cutoff $\hat{\sigma}_{\rm M}$ dependence}

Thus far, we have studied the ED system with the fixed cutoff $\hat{\sigma}_{\rm M}=5$.
Here, we discuss the $\hat{\sigma}_{\rm M}$-dependence of the results.
The system is now characterized by three parameters: the packing fraction $\overline{\phi}$, the width parameter $\gamma$, and the cutoff particle size $\hat{\sigma}_{\rm M}$.
We solved the self-consistent equation for several states ($\overline{\phi}$, $\gamma$, $\hat{\sigma}_{\rm M}$) and found that the critical laws are valid for all $\hat{\sigma}_{\rm M}$ studied.
Thus, we discuss the $\hat{\sigma}_{\rm M}$-dependence of the phase diagram, mainly focusing on the partial glass phase.

To this end, we focus on $\sigma_{\rm M}$-dependence of the critical width parameter $\gamma^{\ast}$ because it controls the presence and location of the partial glass phase. 
We found that $\gamma^{\ast}$ decreases with decreasing $\hat{\sigma}_{\rm M}$, and $\gamma^{\ast}$ becomes $0$ at $\hat{\sigma}_{\rm M} \simeq 2.2$. 
This means that the partial glass phase is absent when $\hat{\sigma}_{M} < 2.2$. 
This tendency is natural because the system approaches the monodisperse system in $\hat{\sigma}_{\rm M} \to 0$. 
On the other hand, we found that $\gamma^{\ast}$ increases with increasing $\hat{\sigma}_{\rm M}$, and $\gamma^{\ast}$ asymptotically approaches $1$ in $\hat{\sigma}_{M}\rightarrow \infty$: $\gamma^{\ast} = 0.94$ for $\hat{\sigma}_{\rm M} = 7$ and $\gamma^{\ast} = 0.99$ for $\hat{\sigma}_{\rm M} = 10$. 
As a result, the phase diagram for larger $\hat{\sigma}_{\rm M}$ is qualitatively similar to Fig.~\ref{exp_cutoff_phase} with $\gamma^{\ast} \simeq 1$.

\section{Summary and Discussion}

In this work, we developed a replica theory of hardspheres with continuous polydispersity in infinite spatial dimensions $d \to \infty$.
We focused on the dynamic glass transition, for which the replica symmetric solution suffices.
We first constructed an equilibrium liquid theory and then extended it to the liquids of replica molecules for continuously polydisperse systems.
By this, we derived the FP potential, an effective free energy describing the glass transition.
In contrast to the monodisperse systems, the cage size $\hat{A}$, the order parameter, depends on the particle size $\hat{\sigma}$ as $\hat{A}(\hat{\sigma})$, and thus, the FP potential is a functional of $\hat{A}(\hat{\sigma})$.
By imposing a stationary point condition, we derived a self-consistent equation for the cage size, Eq.~(\ref{determiningA}).

We first solved Eq.~(\ref{determiningA}) for the nearly monodisperse (NM) system.
The NM system is characterized by a narrow particle size distribution, where a semianalytical solution of Eq.~(\ref{determiningA}) is available.
We found that the critical particle size $\hat{\sigma}^{\ast}$ emerges and that dynamic decoupling occurs; larger particles $\hat{\sigma} \geq \hat{\sigma}^{\ast}$ vitrify, while smaller particles $\hat{\sigma} < \hat{\sigma}^{\ast}$ remain in a liquid state.
In the vicinity of the critical particle size, the cage size exhibits the square-root singularity shown in Eq.~(\ref{NM_cri_size}), which originates from spinodal instability of $\hat{\sigma}^{\ast}$-sized particles as illustrated in Fig.~\ref{F1F2}b.

We next solved Eq.~(\ref{determiningA}) for the exponential distribution (ED) system, which is characterized by more realistic, broader particle size distributions.
As shown in Fig.~\ref{exp_cutoff_phase}, this system has three phases: the liquid, glass, and partial glass phases.
In the partial glass phase, the critical particle size $\hat{\sigma}^{\ast}$ emerges, dynamic decoupling occurs, and the cage size exhibits a square-root singularity at $\hat{\sigma} \simeq \hat{\sigma}^{\ast}$, as in the NM system. 
This means that at any density in the partial glass phase, there are always $\hat{\sigma}^{\ast}$-sized particles in the system, which are on the verge of spinodal instability. 
This is in sharp contrast to the monodisperse system, for which spinodal instability occurs only at the dynamic glass transition density. 

We now compare our results with those of previous works on finite-dimensional systems.
Shimamoto et al.~\cite{shimamoto} numerically and experimentally studied the jamming transition of continuously polydisperse particles with $\phi(\sigma) \propto \sigma^{-a}$ in two-dimensions $d=2$. 
Interestingly, their jamming phase diagram is quite similar to Fig.~\ref{exp_cutoff_phase}, the phase diagram of the ED system. 
They found that the fluid density is maximized at $a = a^{\ast}$ and that many particles become rattlers in the jammed phase at $a < a^{\ast}$. 
Although they studied not the glass transition but the jamming transition, their results are qualitatively similar to our results in that the liquid density is maximized at $\gamma = \gamma^{\ast}$ and the partial glass phase appears at $\gamma < \gamma^{\ast}$. 
We can quantitatively compare $a^{\ast}$ with $\gamma^{\ast}$ because the width parameter $\gamma$ and the exponent $a$ are related as $a = \gamma d$ (see Section IV A). 
Our result $\gamma_{\ast} \simeq 1$ implies $a_{\ast} \simeq d = 2$, which is not far from their observation $2<a_{\ast}<3$.
Extending our theory for the jamming transition would be interesting for a more direct comparison.

The glass transition of power-law distribution systems with $\varphi\left(\sigma\right) \propto \sigma^{-a}$ has been actively studied in simulations, where the standard choice is $a = 3$ in $d=3$. 
Using this system, Pihlajamaa et al.~\cite{Pihlajamaa2023} recently reported that larger particles are caged well while some fraction of smaller particles remain mobile. 
This observation is qualitatively similar to the phenomenology of the partial glass phase in the ED system. 
As we noted, our result $\gamma_{\ast} \simeq 1$ implies that the partial glass phase emerges for $a \leq a_{\ast} \simeq 3$. 
It would be interesting to systematically study the glass transition for $a < 3$ in simulations, which will provide more information on the partial glass phase. 

Finally, we mention two additional research directions to extend our work.
First, it is interesting to study the Gardner transition of continuously polydisperse hardspheres. 
As we showed, this system has a partial glass phase, in which $\hat{\sigma}^{\ast}$-sized particles are on the verge of spinodal instability. 
It is highly nontrivial how the Gardner transition occurs by compressing such partial glasses. 
It is also interesting to extend our theory to finite-dimensional systems by combining replica liquid theory with finite-dimensional liquid state theory.
This should enable us to compare the theory with simulations and experiments more quantitatively.

\begin{acknowledgments}
We thank H.~Ikeda, Y.~Tomita, H.~Mizuno, F.~Ghimenti, and F.~van Wijland for insightful discussions and useful exchanges. 
This work was supported by JSPS KAKENHI Grant Numbers JP20H01868 and JP20H00128.
\end{acknowledgments}

\bibliography{poly}

\end{document}